%% file: Paper2563.tex
\documentclass[runningheads]{llncs}
\usepackage{graphicx}
\usepackage{amsmath}
\usepackage{subfiles}
\begin{document}
\title{Deep Learning-based Universal Beamformer for Ultrasound Imaging\thanks{This work is supported by National Research Foundation of Korea, Grant Number: NRF-2016R1A2B3008104.}}
%
%
\author{Shujaat Khan \inst{1}\orcidID{0000-0001-9676-6817} \and
Jaeyoung Huh\inst{1}\orcidID{0000-0002-2126-0763} \and
Jong Chul Ye\inst{1}\orcidID{0000-0001-9763-9609}}

\institute{Department of Bio and Brain Engineering, Korea Advanced Institute of Science and Technology (KAIST), 335 Gwahangno, Yuseong-gu, Daejeon 305-701, Korea.\\
\email{\{shujaat,woori93,jong.ye\}@kaist.ac.kr}}
\maketitle              
\begin{abstract}
In ultrasound (US) imaging, individual channel RF measurements are back-propagated and accumulated to form an image after applying specific delays. While this time reversal is usually implemented using a hardware- or software-based delay-and-sum (DAS) beamformer, the performance of DAS decreases rapidly in situations where data acquisition is not ideal.  
 Herein, for the first time, we demonstrate that a single data-driven adaptive beamformer designed as a deep neural network
 can generate high quality images robustly for various detector channel configurations and subsampling rates.  
The proposed deep beamformer is evaluated
 for two distinct acquisition schemes:  focused ultrasound imaging and  planewave imaging. Experimental
 results showed that the proposed deep beamformer exhibit significant performance gain  for both focused and planar
 imaging schemes, in terms of contrast-to-noise ratio and structural similarity.

\end{abstract}

\section{Introduction}
Due to minimal invasiveness from non-ionizing radiations and excellent temporal resolution, ultrasound (US) imaging is an indispensable tool for various clinical applications such as cardiac, fetal imaging, etc.  The basic imaging principle of US imaging is based on the time-reversal  \cite{TimeReversal1,TimeReversal2}, which is based on a mathematical observation that the wave operator is self-adjoint. For example, in focused B-mode US imaging, the return echoes from individual scan-line are recorded by the receiver channels,
after which delay-and-sum (DAS) beamformer applies the time-reversal delay to the channel measurement and additively combines them for each time point to form images at each scan-line. {Although  sonic inhomogeneities can cause wavefront aberrations in ultrafast imaging \cite{4012867},
 the estimation of the time delay in the DAS beamformer is based primarily on the assumption of a constant sound speed. }
Due to the simplicity, high-speed analog-to-digital converters (ADCs) and large number of receiver elements are often necessary in to reduce side lobes which otherwise degrade image resolution and contrast. To address this problem, various adaptive beamforming techniques have been developed over the several decades \cite{AdaptiveBF1,MVBF1,IterativeBF1}.

Recently, inspired by the tremendous success of deep learning, the authors in \cite{yoon2018efficient,gasse2017high} use deep neural networks for the reconstruction of high-quality US images from limited number of received {radio frequency (RF)} data.   
For example, the work in \cite{gasse2017high} uses deep neural network for coherent compound imaging from small number of plane wave illumination. In focused B-mode ultrasound imaging, \cite{yoon2018efficient} employs the deep neural network to interpolate the missing RF-channel data with multiline aquisition for accelerated scanning. While these recent deep neural network approaches provide impressive reconstruction performance,  the current design is not universal in the sense that they are designed and trained for specific acquisition scenario.

Therefore, one of the most important contributions of this paper is to demonstrate that a single end-to-end  beamformer implemented by a deep neural network
(DeepBF) can generate high quality images robustly for various detector channel configurations and subsampling rates.  The main innovation of our universal deep beamformer comes from one of the most exciting properties of deep neural network - exponentially increasing expressiveness with respect to the channel and depth \cite{ye2019understanding}. Thanks to the expressiveness of  neural networks, our DeepBF can learn large number of
 mappings between
various cases of RF measurements and images, and exhibits superior image quality for all sub-sampling rates. 
Moreover, unlike \cite{yoon2018efficient} that interpolates the missing RF data and uses standard DAS beamformer,  the novelty of this work is the end-to-end deep learning to replace the standard BF, which was never considered in \cite{yoon2018efficient}. Consequently, our approach is much simpler and can be easily incorporated to replace the standard beamforming pipeline. Inspite of the simplicity, we will show that this approach outperforms
the interpolation approach in \cite{yoon2018efficient}.
Another important byproduct of the proposed method is that the trained neural network can  further  improve the image contrast even for the full rate cases.
The origin of this performance enhancement is also analyzed.


\section{Method}
\label{sec:methods}

\subsection{Dataset}

Multiple RF data were acquired using a linear array transducer (L3-12H) with a center frequency of $8.48$ MHz on E-CUBE 12R US system (Alpinion Co., Korea). The configuration of the probe is given in Table~\ref{probe_config}. 

\begin{table}[!hbt]
	\centering
	\caption{Probe Configuration}
	\label{probe_config}
	\resizebox{0.5\textwidth}{!}{
		\begin{tabular}{c|c}
			\hline
			{Parameter} & {Linear Probe} \\ \hline\hline
			Probe Model No.& L3-12H \\
			Carrier wave frequency & 8.48 MHz \\
			Sampling frequency & 40 MHz \\
			No. of probe elements & 192 \\
			No. of Tx elements & 128 \\
			No. of TE events (focused mode) & 96\\
			No. of Rx elements (focused/unfocused mode) & 64/192\\
			No. of PWs (unfocused mode) & 31\\
			Elements pitch & 0.2 mm\\
			Elements width & 0.14 mm \\
			Elevating length & 4.5 mm\\  \hline
	\end{tabular}
}
\end{table}

Using a linear probe, we acquired RF data from the carotid area of $10$ volunteers.  In focused mode imaging experiment the \textit{in-vivo} data consists of $40$ temporal frames per subject, providing $400$ sets of {Depth-Receiver channels-transmit events (Depth-Rx-TE)} data cube.  The dimension of each Rx-TE plane was $64\times96$.  A set of $30,000$ Rx-TE planes was randomly selected from the $4$ subjects datasets, and data cubes (Depth-Rx-TE) are then divided into $25,000$ datasets for training and $5000$ datasets for validation.  The remaining dataset of $360$ frames was used as a test dataset. 

In plane wave imaging experiments,  we acquire $109$ frames, among which only $8$ frames (images) from in-vivo data were used for training and $1$ for validation purpose while remaining $100$ were used as test dataset. Each US image raw data consist of $31$ PWs and $192$-channels, and each frame have different depth ranges varying from $25-60$ mm consist of $2000-9000$ depth planes. 

{For quantitative evaluation, we also acquired RF data from the ATS-539 multipurpose tissue mimicking phantom. These datasets were only used for test purpose and no additional training of CNN was performed on it.  The phantom datasets were used to verify the generalization power of the proposed method.}

\subsection{RF sub-sampling scheme}
\label{sec:RFsamScon}
For focused mode imaging, we generated six sets of sub-sampled RF data at different down-sampling rates. In addition to the full RF data (i.e. $64$-channels),
 we use several sub-sampling cases using $32$, $24$, $16$, $8$ and $4$ Rx-channels. Since the active receivers at the center of the scan-line get RF data from direct reflection, two channels that are in the center of active transmitting channels were always included to improve the performance, and remaining channels were randomly selected from the receiving channels. For each depth plane, a different sampling pattern (mask) is used. The non-active Rx channels are zero-padded.

For unfocused planar wave imaging,  in addition to the full RF data, we generated six sets of sub-sampled RF data at different down-sampling rates. In particular, we used two subsampling schemes: variable down-sampling of RF-channel data pattern across the depth to reduce high data-rate and power requirements, and uniform sub-sampling of PWs angles to accelerate acquisition speed.  Here we use the following subsampling cases: (1) $64$, $32$, $16$, and $8$ Rx-channels with $31$ PWs. (2) $31$, $11$, $7$, and $3$ PWs with $64$ Rx-channels.

\subsection{Network architecture}
\begin{figure}
	\centering{\includegraphics[width=10cm]{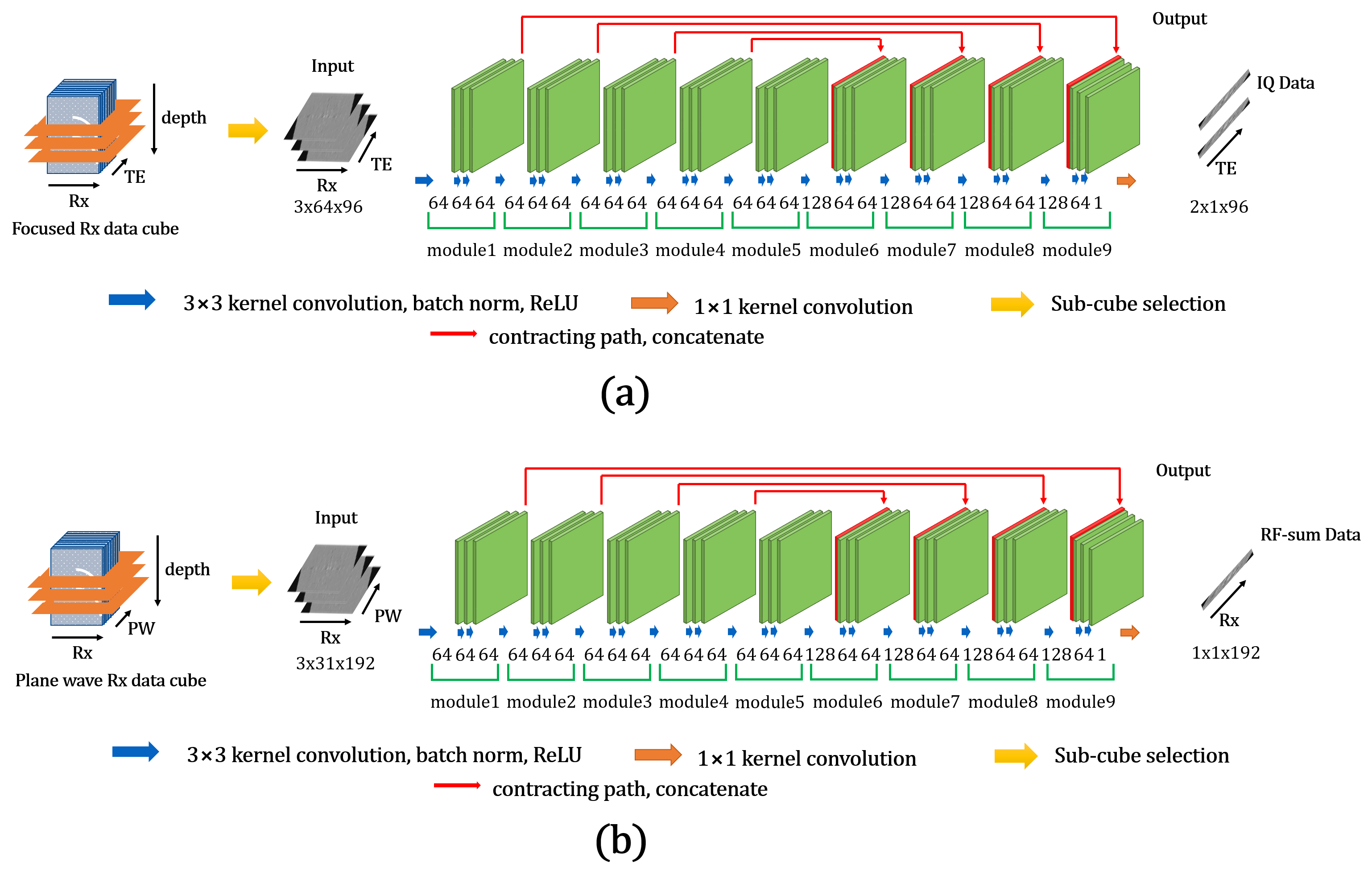}}
	\caption{Proposed CNN architecture for sub-sampled (a) focused US B-mode imaging. (b) planewave US B-mode imaging.} \label{fig:CNN_block_diagram}
\end{figure}

{{In focused mode imaging}, $3\times64\times96$ data-cube in the Depth-Rx-TE sub-space was used for CNN training to generate a $2\times3\times96$ I and Q data in the Depth-TE plane. Here IQ data is the Hilbert transformed data before envelope detection. The input cube consist of three adjacent depth planes and the target IQ data of middle plane is obtained from two output channels each representing real and imaginary parts. The proposed CNN consists of $27$ convolution layers composed of a contracting path with concatenation, batch normalization, ReLU except for the last convolution layer. The first $26$ convolution layers use $3\times3$ convolutional filters (i.e., the 2-D filter has a dimension of $3\times 3$), and the last convolution layer uses a $1\times1$ filter and contract the $3\times64\times96$ data-cube from Depth-Rx-TE sub-space to $2\times1\times96$ IQ-Depth-TE plane as shown in Figs.~\ref{fig:CNN_block_diagram}(a). 
}

In plane wave imaging
 a multi-channel CNN was trained using $3\times31\times192$ data-cube in the Depth-PW-Rx sub-space to generate a $1\times192$ RF sum data in the Depth-TE plane. Three input channels were used to process three adjacent depth planes to generate target RF sum data of the central depth plane.  The proposed CNN consists of $27$ convolution layers composed of a contracting path with concatenation, batch normalization, and ReLU except for the last convolution layer. The first $26$ convolution layers use $3\times3$ convolutional filters (i.e., the 2-D filter has a dimension of $3\times 3$), and the last convolution layer uses a $1\times1$ filter and contract the $3\times31\times192$ data-cube from Depth-PW-Rx sub-space to $1\times1\times192$ Depth-Rx plane as shown in Figs.~\ref{fig:CNN_block_diagram}(b). 

Both networks were implemented with MatConvNet \cite{vedaldi2015matconvnet} in the MATLAB 2015b environment. Specifically, for network training, the parameters were estimated by minimizing the $l_2$ norm loss function using a stochastic gradient descent with a regularization parameter of $10^{-4}$. The learning rate started from $10^{-3}$ and gradually decreased to $10^{-5}$ in $200$ epochs. The weights were initialized using Gaussian random distribution with the Xavier method \cite{glorot2010understanding}. 

\section{Experimental Results}
\label{sec:results}
To quantitatively show the advantages of the proposed deep learning method, we used the contrast-to-noise ratio (CNR), generalized CNR (GCNR) \cite{GCNR_Paper}, and structure similarity (SSIM).
\subsubsection{Focused mode imaging}
\begin{figure*}[!hbt]
	\centerline{\includegraphics[width= 10cm]{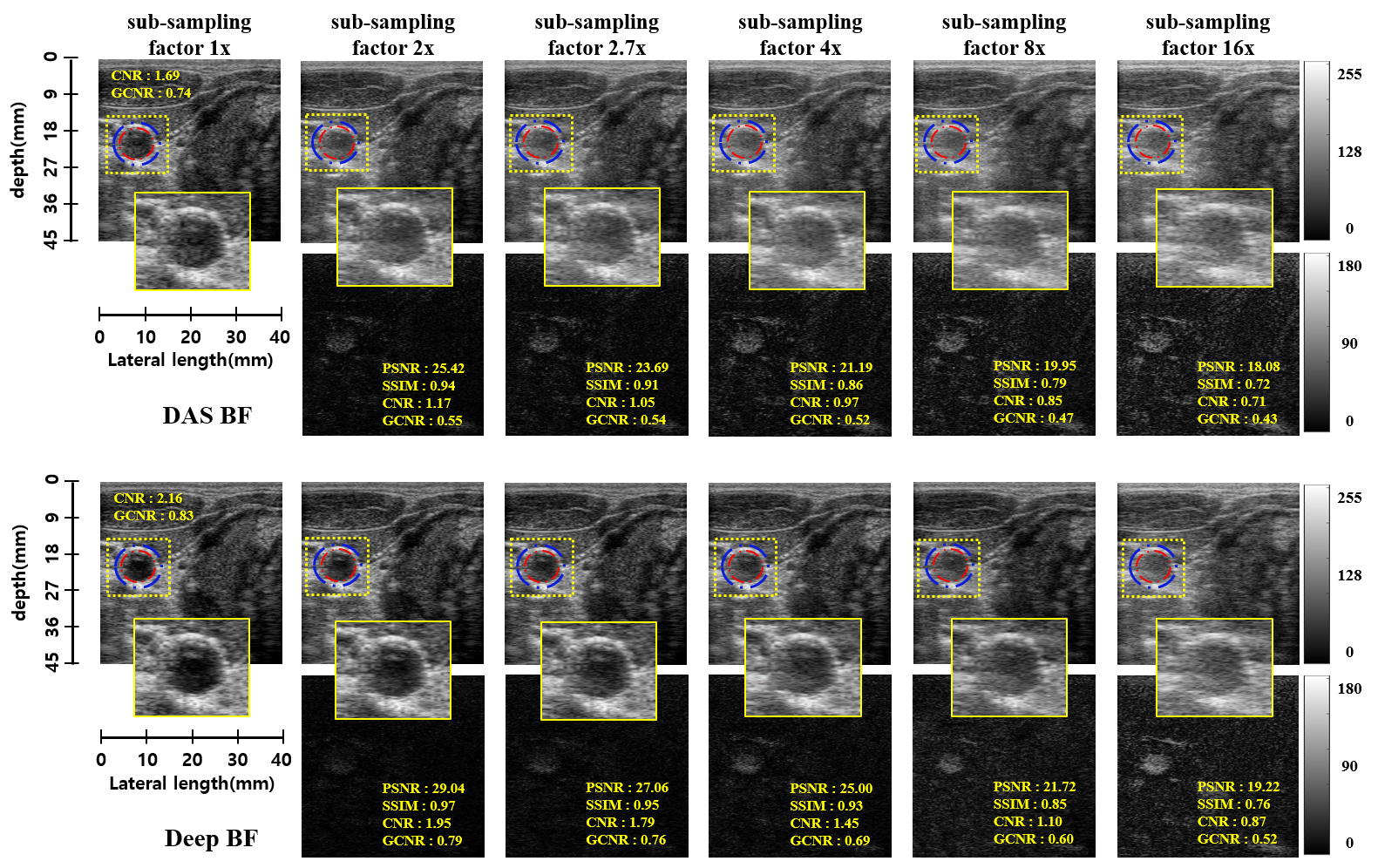}}
	\caption{\footnotesize Focused B-mode imaging reconstruction results of standard DAS beam-former and the proposed method for in-vivo carotid region.}
	\label{fig:results_view_focused}	
\end{figure*} 

Figs.~\ref{fig:results_view_focused}  show the results of an \textit{in vivo} example for the full data (i.e. $64$ channel) as well as $32$, $24$, $16$, $8$ and $4$ Rx-channels down-sampling schemes.  Since 64 channels are used as a full sampled data, this corresponds to $1\times, 2\times, 2.7 \times, 4\times, 8\times$ and $16\times$ sub-sampling factors.  The images are generated using the proposed DeepBF and the standard DAS beam-former method.   Our method significantly improves the visual quality of the US images by estimating the correct dynamic range and eliminating artifacts for both sampling schemes.  From difference images, it is evident that the quality degradation of images in DAS is higher than the DeepBF. Note that the proposed method successfully reconstruct both the near and the far field regions with equal efficacy, and only minor structural details are imperceivable. Furthermore, it is remarkable that the CNR and GCNR values are significantly improved by the DeepBF even for the fully sampled case (eg. from $1.69$ to $2.16$ in CNR and from $0.74$ to $0.83$ in GCNR), which clearly shows the advantages of the proposed method. 
It is believed that this is due to performance boosting behavior similar to super-resolution deep neural networks in which the authors have found that multiple-magnification training outperforms fixed-magnification cases \cite{kim2016accurate}.
Additional results on phantom dataset are available in supplemental document.
\begin{table}[!hbt]
	\centering
	\caption{Focus B-mode imaging performance comparison for  \textit{in vivo} data using variable sampling pattern}
	\label{tbl:results_vSTATS_invivo}
	\resizebox{0.7\textwidth}{!}{
		\begin{tabular}{c|cccccccc}
			\hline
			\textbf{sub-sampling} & \multicolumn{2}{c}{\textbf{CNR}} & \multicolumn{2}{c}{\textbf{GCNR}} & \multicolumn{2}{c}{\textbf{PSNR (dB)}} & \multicolumn{2}{c}{\textbf{SSIM}}  \\
			\textbf{factor} & \textit{DAS} & \textit{DeepBF} & \textit{DAS} & \textit{DeepBF} & \textit{DAS} & \textit{DeepBF} & \textit{DAS} & \textit{DeepBF}  \\ \hline\hline
			1 & 1.38 & 1.45 & 0.64 & 0.66 & $\infty$ & $\infty$ & 1 & 1 \\
			2 & 1.33 & 1.47 & 0.63 & 0.66 & 24.59 & 27.38 & 0.89 & 0.95 \\
			2.7 & 1.3 & 1.44 & 0.62 & 0.66 & 23.15 & 25.54 & 0.86 & 0.92 \\
			4 & 1.25 & 1.38 & 0.6 & 0.64 & 21.68 & 23.55 & 0.81 & 0.87 \\
			8 & 1.18 & 1.26 & 0.58 & 0.6 & 19.99 & 21.03 & 0.74 & 0.77  \\
			16 & 1.12 & 1.17 & 0.56 & 0.58 & 18.64 & 19.22 & 0.67 & 0.69 \\  \hline
		\end{tabular}
	}
\end{table}


We also compared the CNR, GCNR, PSNR, and SSIM distributions of reconstructed B-mode images obtained from $360$ \textit{in-vivo}  test frames. Table~\ref{tbl:results_vSTATS_invivo} showed that the proposed deep beamformer consistently outperformed the standard DAS beamformer for all subsampling schemes and ratios.  One big advantage of ultrasound image modality is it run-time imaging capability, which require fast reconstruction time. Another important advantage of the proposed method is the run-time complexity. 
The average reconstruction time for each depth planes is around $4.8$ (milliseconds), which could be easily reduce by optimized implementation and reconstruction of multiple depth planes in parallel.

{In contrast to \cite{yoon2018efficient} where deep learning approach was designed for interpolating missing RF data to be used as input for standard beamformer (BF), the proposed method is an end-to-end CNN-based beamforming pipeline, without requiring additional BF.   To verify that the new approach does not sacrifice any performance, we performed quantitative study using test dataset.
As shown in Supplementary Material, the proposed network outperform the method in \cite{yoon2018efficient}  in both CNR and GCNR. }

\begin{figure}[!hbt]
	\centerline{\includegraphics[width= 10cm]{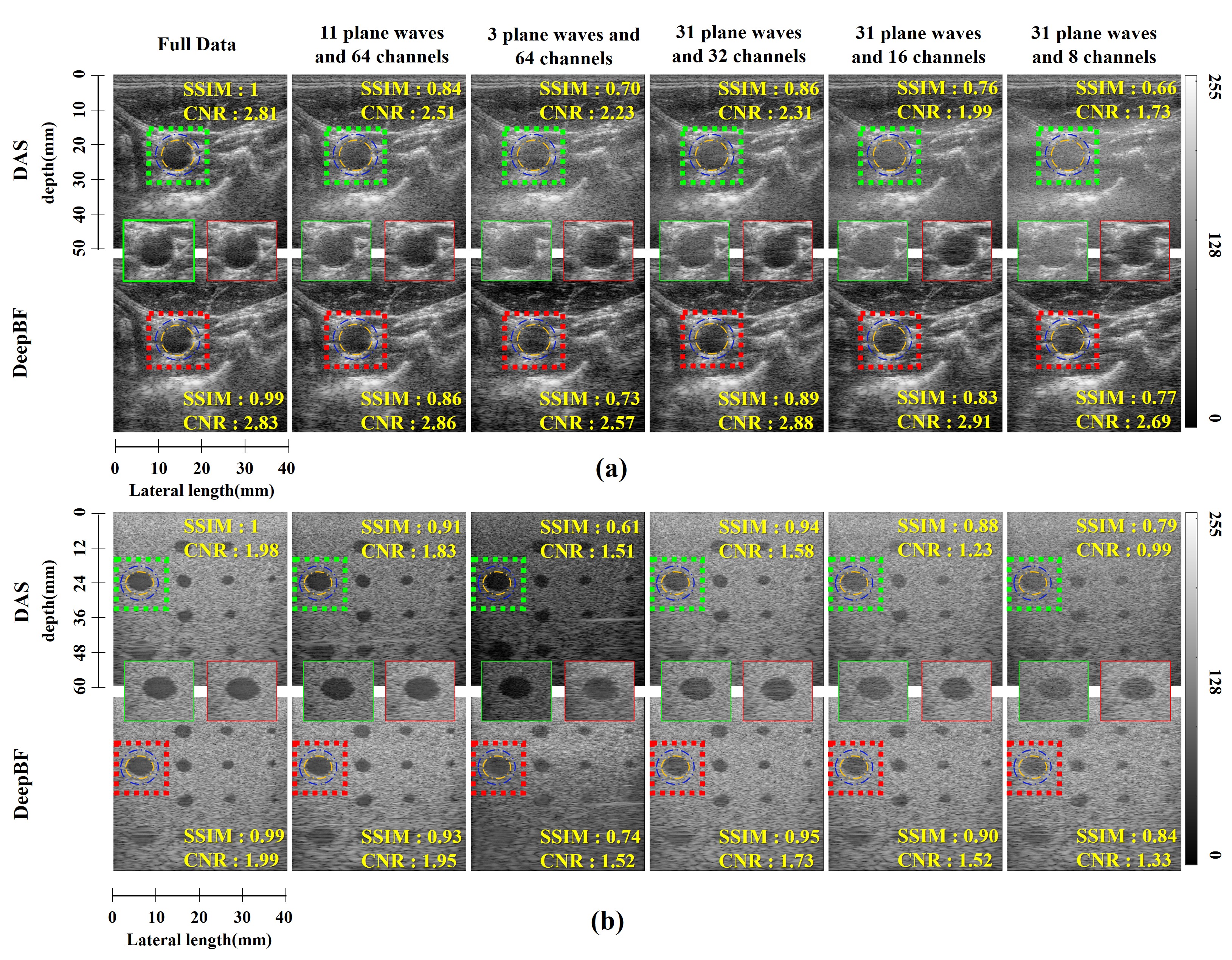}}
	\caption{\footnotesize Planewave B-mode imaging reconstruction results of standard DAS beam-former and the proposed method for: (a) in-vivo carotid region (b) tissue mimicking phantom.}
	\label{fig:results_view_PW}	
\end{figure} 
\subsubsection{Planewave US imaging}
Figs.~\ref{fig:results_view_PW}(a)(b) show the results \textit{in vivo} and phantom image examples for different down-sampling schemes.  The images are generated using the proposed DeepBF and the standard DAS beam-former method.  Our method significantly improves the visual quality of the US images by estimating the correct dynamic range and eliminating  artifacts for both sampling schemes.  From zoomed region images, it can be seen that the quality of the DeepBF images is relatively unchanged for variable sampling scenarios. Note that the proposed method successfully reconstruction both the near and the far field regions with equal efficacy, and only minor structural details are imperceivable. 
\begin{figure}[!hbt]
	\centerline{\includegraphics[width=5cm]{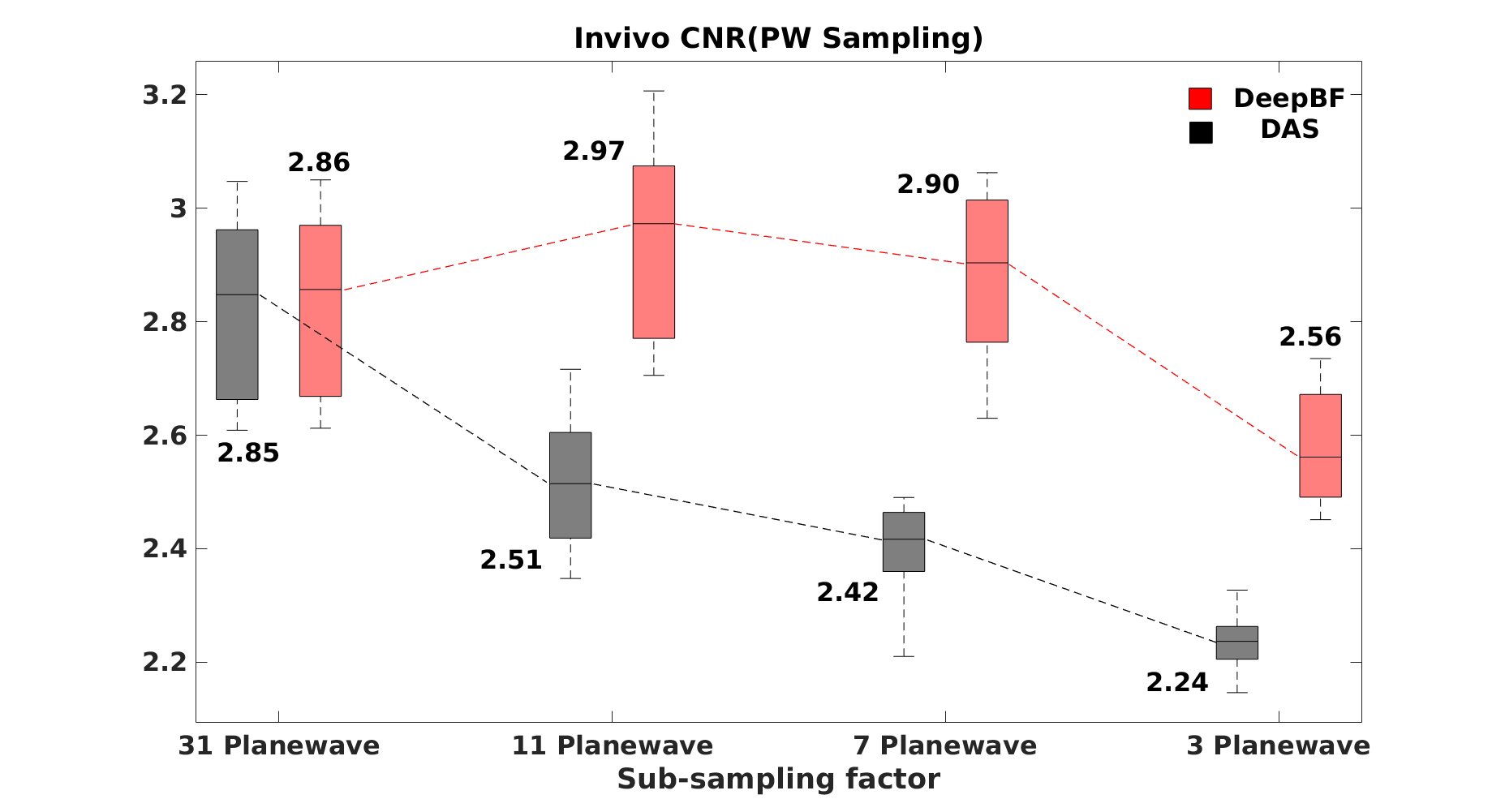}\hspace*{-0.5cm}\includegraphics[width=5cm]{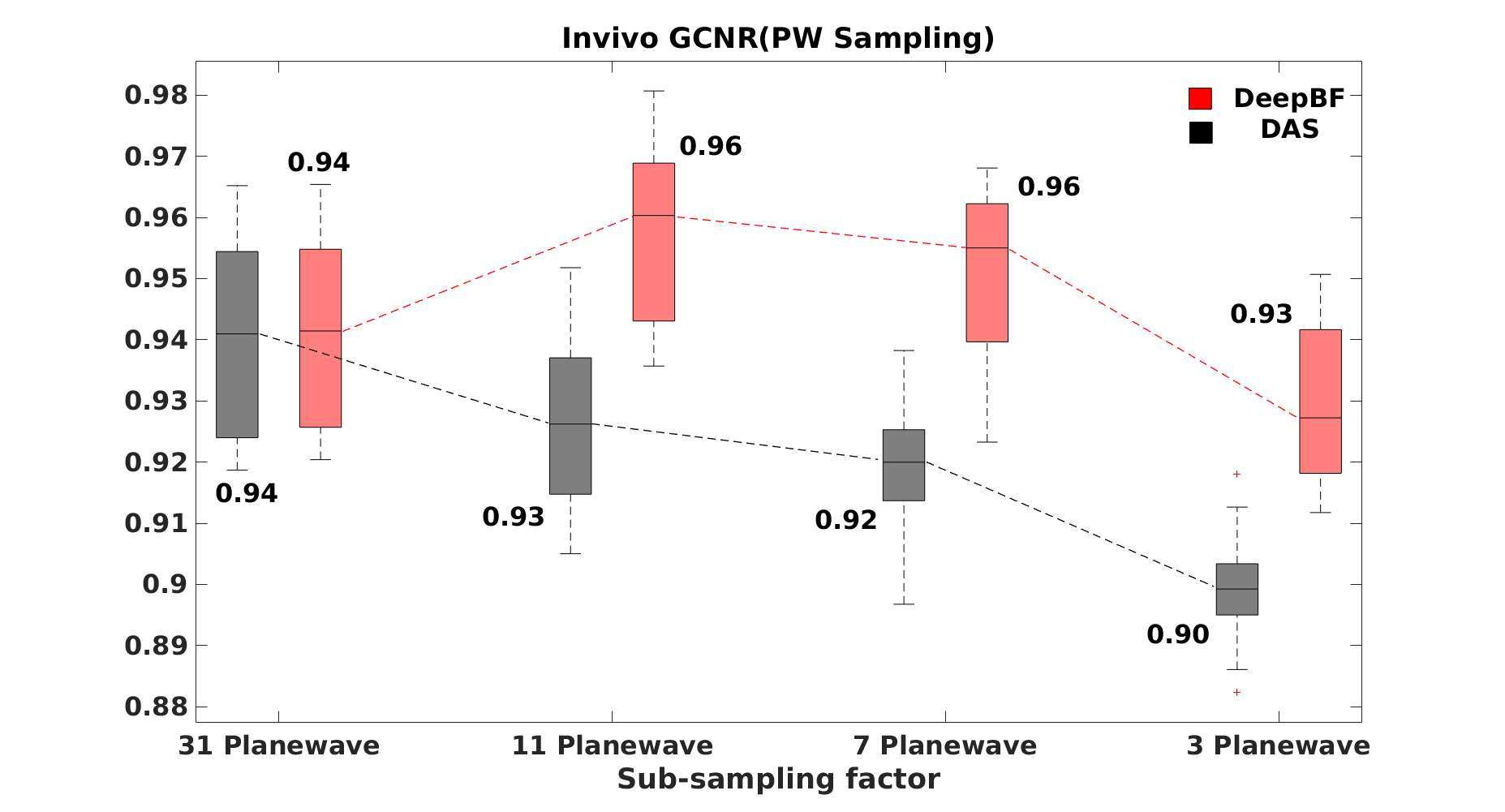}\hspace*{-0.5cm}\includegraphics[width=5cm]{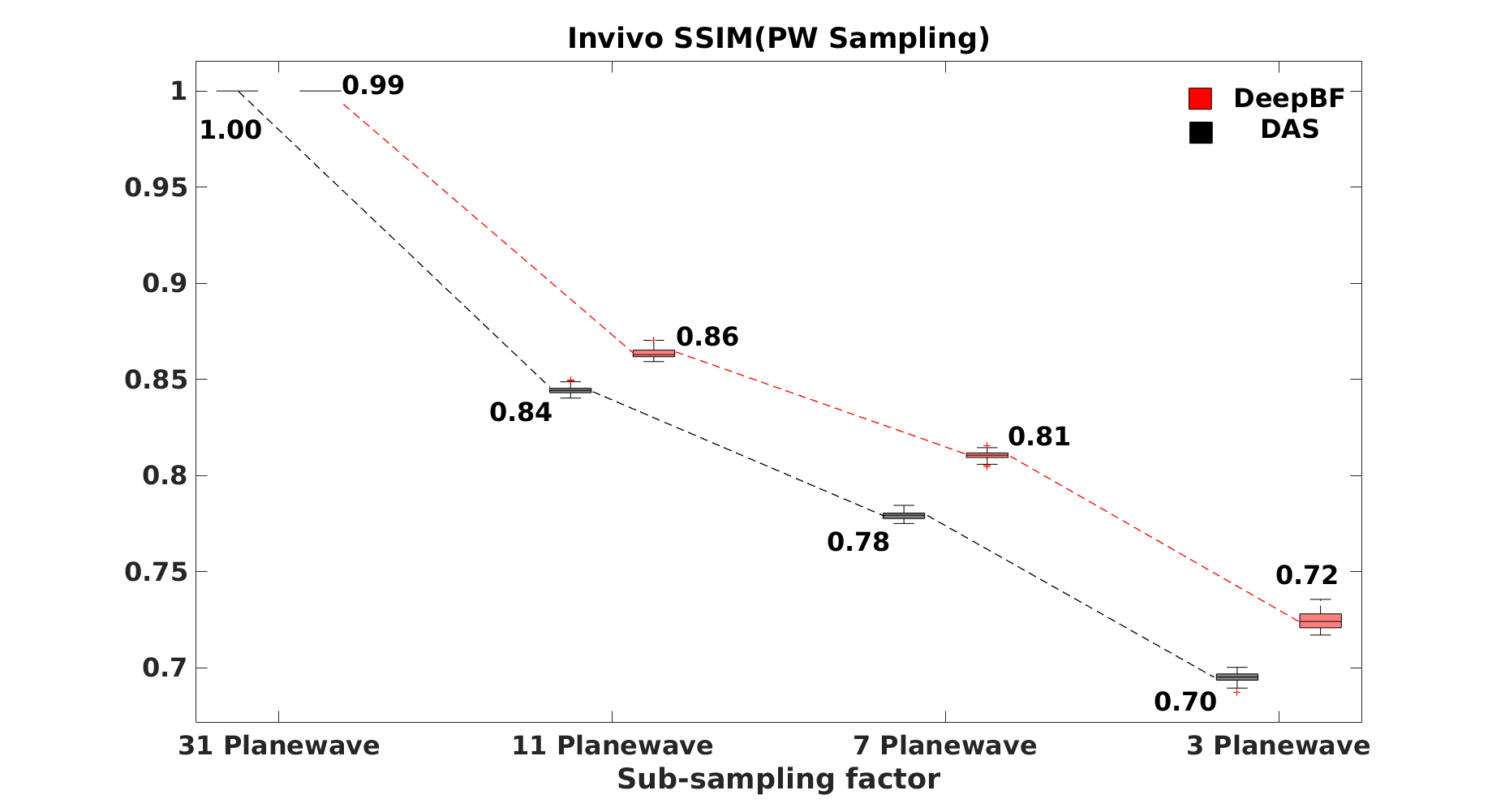}\hspace*{-0.5cm}}
	\vspace*{0.1cm}
	\caption{\footnotesize Quantitative comparison using invivo data on different subsampling schemes in plane wave imaging: ({first column}) CNR value distribution, ({second column}) GCNR value distribution, ({third column}) SSIM value distribution. }
	\label{fig:results_STATS_PW}
\end{figure}
In addition, we compared the CNR, GCNR, and SSIM distributions of reconstructed B-mode images obtained from $100$ {invivo} test frames. Our method shows significant performance gain in all measures.  From Fig.~\ref{fig:results_STATS_PW}, it is evident that the quality degradation of images in DAS is higher than the DeepBF. 
Furthermore, it is remarkable that the CNR value are significantly improved by the DeepBF even for the fully sampled case (eg. from $2.85$ to $2.86$ in CNR), which clearly shows the advantages of the proposed method.

\section{Conclusion}
\label{sec:conclusion}

Herein, for the first time we demonstrated that a single {deep beamformer} trained using a deep neural network way can be used for variable rate ultrasound imaging. Even for fully sampled data, the proposed method further improves the images. 
Moreover, CNR, GCNR, PSNR, and SSIM were significantly improved over standard DAS method across various subsampling schemes. 
{The proposed schemes may substantially help in designing low-powered accelerated ultrasound imaging systems. Runtime implementation for cardiac or fetus imaging may help in investigating clinical significance of the proposed method.}

%
%
%
 \bibliographystyle{splncs04}
\bibliography{Paper2563}
%
%
%
%
%

\subfile{Paper2563_Supplemental.tex}
\end{document}

%% file: Paper2563_Supplemental.tex
%
\title{--- Supplemental Document --- \\Deep Learning-based Universal Beamformer for Ultrasound Imaging\thanks{This work is supported by National Research Foundation of Korea, Grant Number: NRF-2016R1A2B3008104.}}
%
%
\author{Shujaat Khan \inst{1}\orcidID{0000-0001-9676-6817} \and
Jaeyoung Huh\inst{1}\orcidID{0000-0002-2126-0763} \and
Jong Chul Ye\inst{1}\orcidID{0000-0001-9763-9609}}

\institute{Department of Bio and Brain Engineering, Korea Advanced Institute of Science and Technology (KAIST), 335 Gwahangno, Yuseong-gu, Daejeon 305-701, Korea.\\
\email{\{shujaat,woori93,jong.ye\}@kaist.ac.kr}\\
\url{https://bispl.weebly.com}}
\maketitle              
%

%
%
%
%
\section{Experimental Results}
\label{sec:results}
\subsubsection{Focused mode imaging results on phantom dataset}

\begin{figure*}[!hbt]
	\centerline{\includegraphics[width= \textwidth]{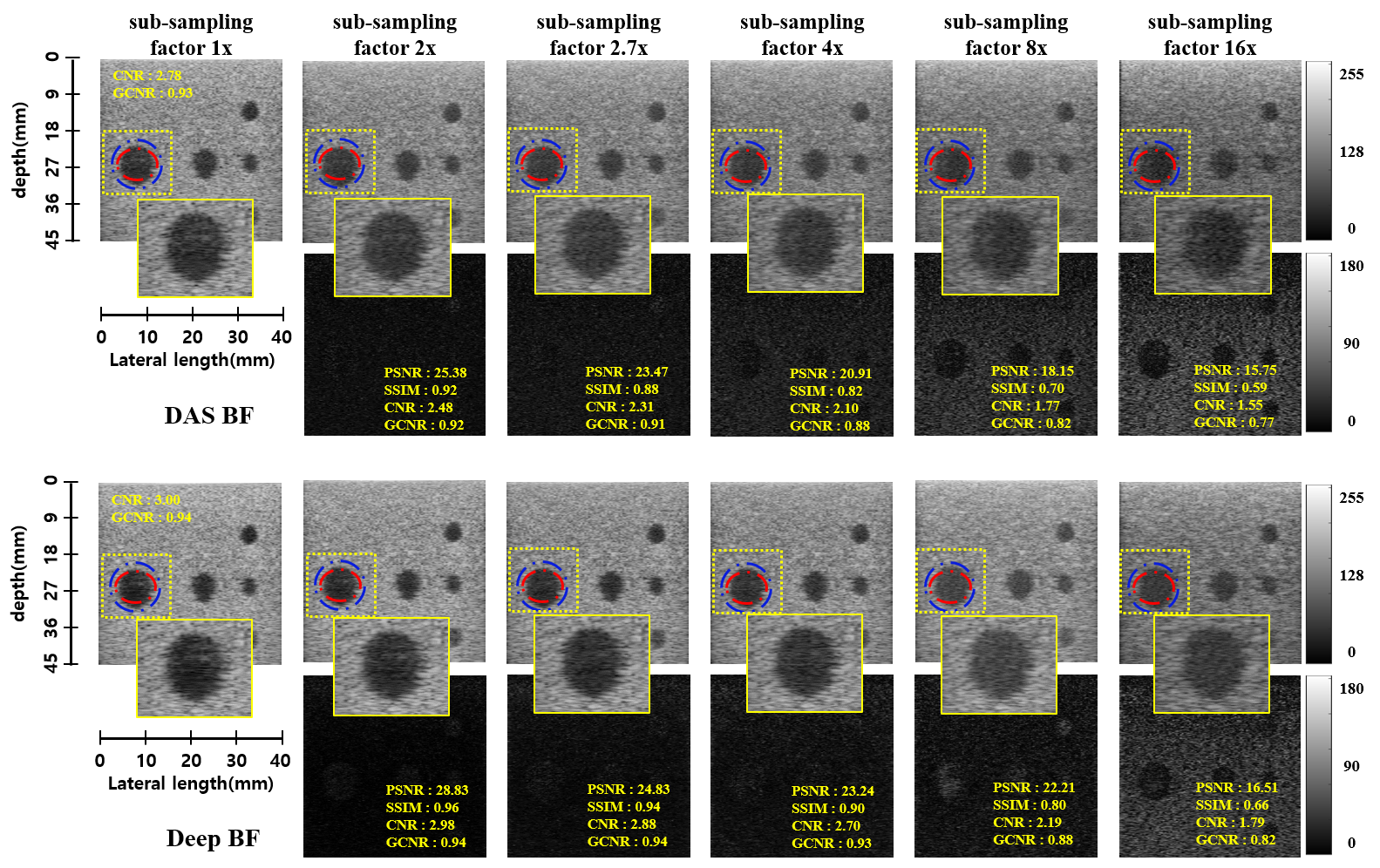}}
	\caption{\footnotesize Focused B-mode imaging reconstruction results of standard DAS beam-former and the proposed method for anechoic phantom cyst. All images are shown with a $60$ dB dynamic range.}
	\label{fig:results_view_focused}	
\end{figure*}

Fig \ref{fig:results_view_focused} shows the reconstruction results on phantom dataset.  We also compared the CNR, GCNR, PSNR, and SSIM distributions of reconstructed B-mode images obtained from $159$ \textit{phantom}  test frames. Table~\ref{tbl:results_vSTATS_phantom} showed that the proposed deep beamformer consistently outperformed the standard DAS beamformer for all subsampling schemes and ratios.  

\begin{table}[!hbt]
	\centering
	\caption{Performance statistics on \textit{phantom} data for variable sampling pattern using focused mode imaging}
	\label{tbl:results_vSTATS_phantom}
	\resizebox{0.8\textwidth}{!}{
		\begin{tabular}{c|cccccccc}
			\hline
			\textbf{sub-sampling} & \multicolumn{2}{c}{\textbf{CNR}} & \multicolumn{2}{c}{\textbf{GCNR}} & \multicolumn{2}{c}{\textbf{PSNR (dB)}} & \multicolumn{2}{c}{\textbf{SSIM}}  \\
			\textbf{factor} & \textit{DAS} & \textit{DeepBF} & \textit{DAS} & \textit{DeepBF} & \textit{DAS} & \textit{DeepBF} & \textit{DAS} & \textit{DeepBF}  \\ \hline\hline
			1 & 2.59 & 2.66 & 0.897 & 0.897 & $\infty$ & $\infty$ & 1 & 1 \\
			2 & 2.47 & 2.65 & 0.896 & 0.900 & 25.51 & 28.24 & 0.91 & 0.95 \\
			2.7 & 2.36 & 2.62 & 890 & 0.901 & 23.51 & 26.57 & 0.87 & 0.92 \\
			4 & 2.20 & 2.53 & 0.875 & 0.898 & 21.44 & 24.55 & 0.81 & 0.88 \\
			8 & 1.93 & 2.25 & 0.839 & 0.874 & 18.97 & 21.27 & 0.71 & 0.78  \\
			16 & 1.68 & 1.92 & 0.789 & 0.828 & 17.20 & 18.51 & 0.60 & 0.66 \\  \hline
		\end{tabular}
	}
\end{table}

\subsubsection{Comparison with deep RF interpolation \cite{yoon2018efficient}}
In \cite{yoon2018efficient}, deep learning approach was designed for interpolating missing RF data, which are later used as input for standard beamformer (BF). On the other hand, the proposed method is an end-to-end CNN-based beamforming pipeline, without requiring additional BF.  Consequently, our approach is much simpler and can be easily incorporated to replace the standard beamforming pipeline. 

\begin{figure}[!hbt]
	\centerline{\includegraphics[width= \textwidth]{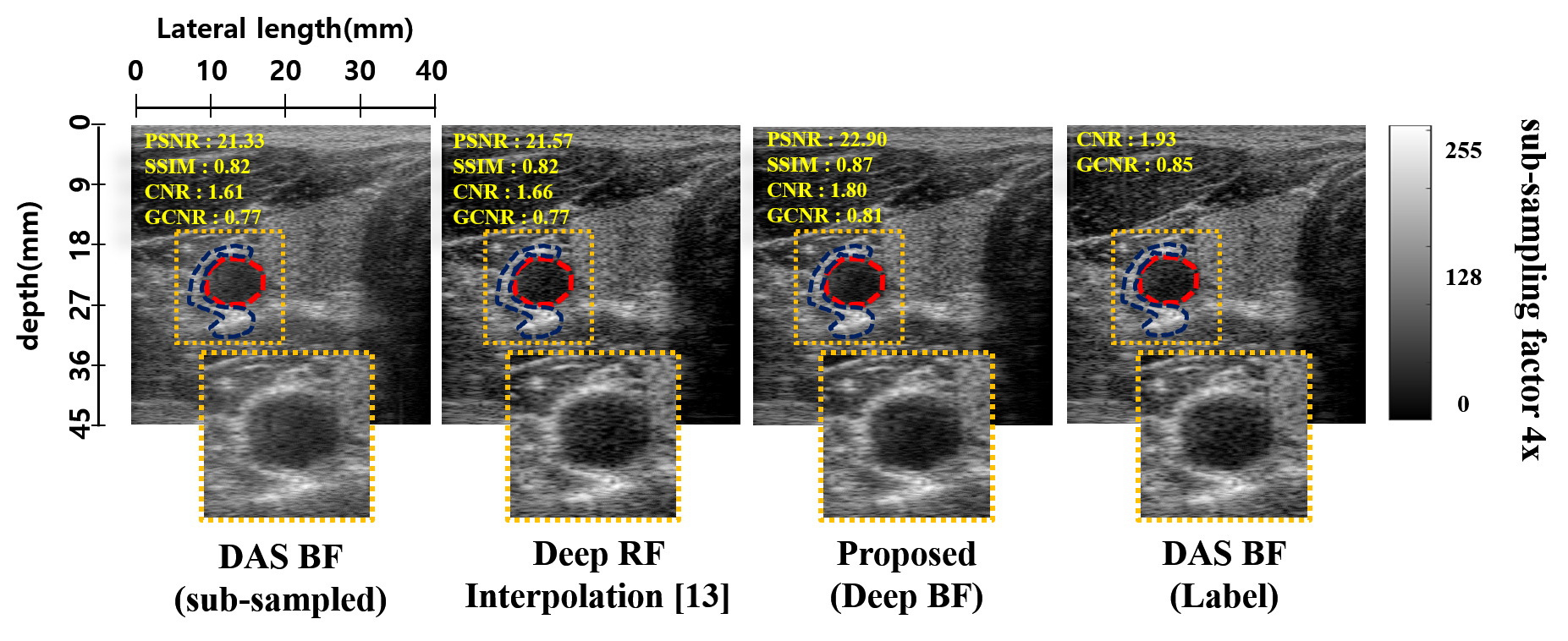}}
	\caption{\footnotesize Focused B-mode imaging reconstruction results of standard DAS beam-former, Deep RF Interpolation \cite{yoon2018efficient} and the proposed DeepBF for 4$\times$ sub-sampled \textit{in-vivo} data. All images are shown with a $60$ dB dynamic range.}
	\label{fig:results_view_Efficient}	
\end{figure}

\begin{figure}[!hbt]
	\centerline{\includegraphics[width=4cm]{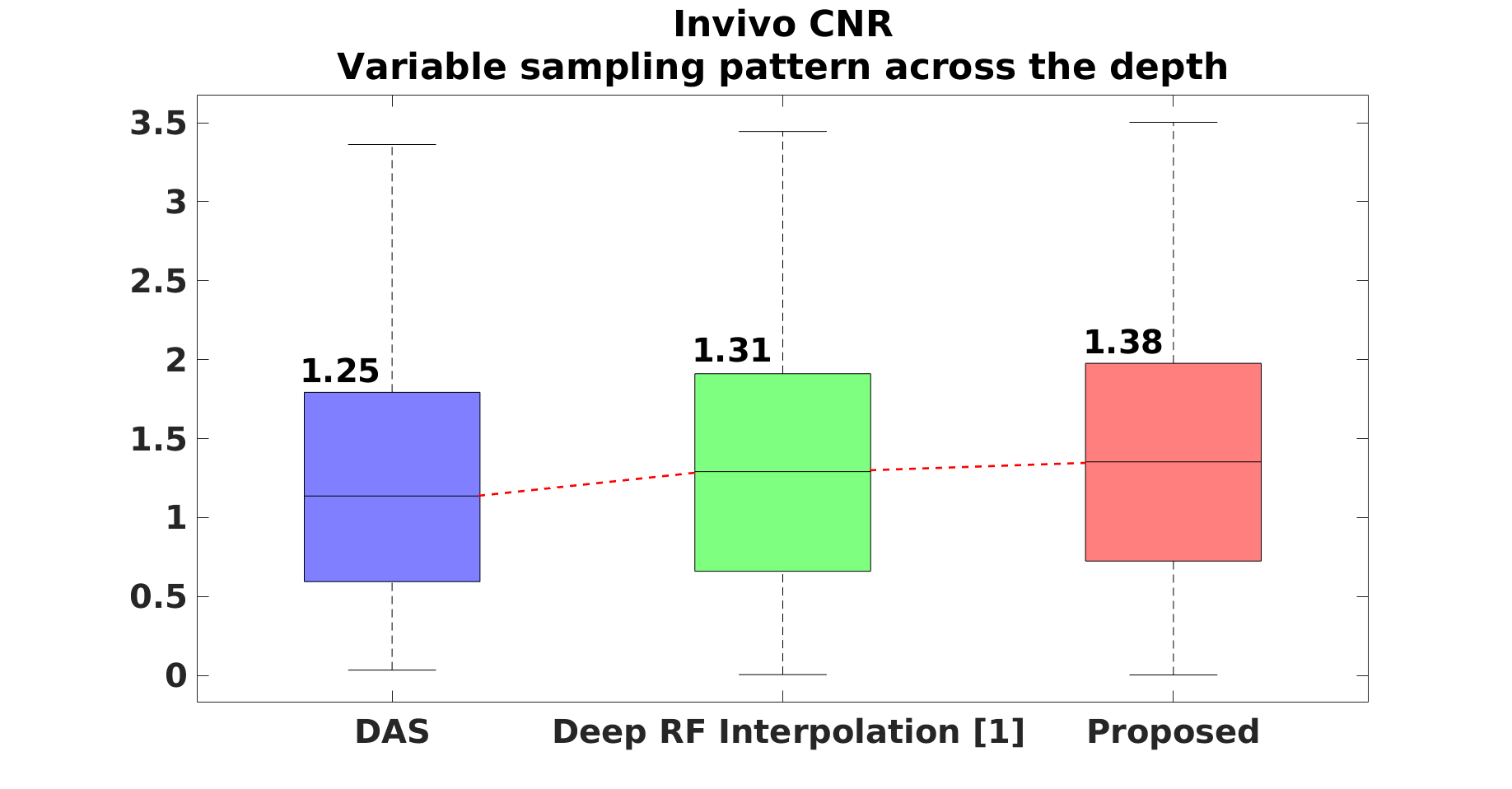}\hspace*{-0.5cm}\includegraphics[width=4cm]{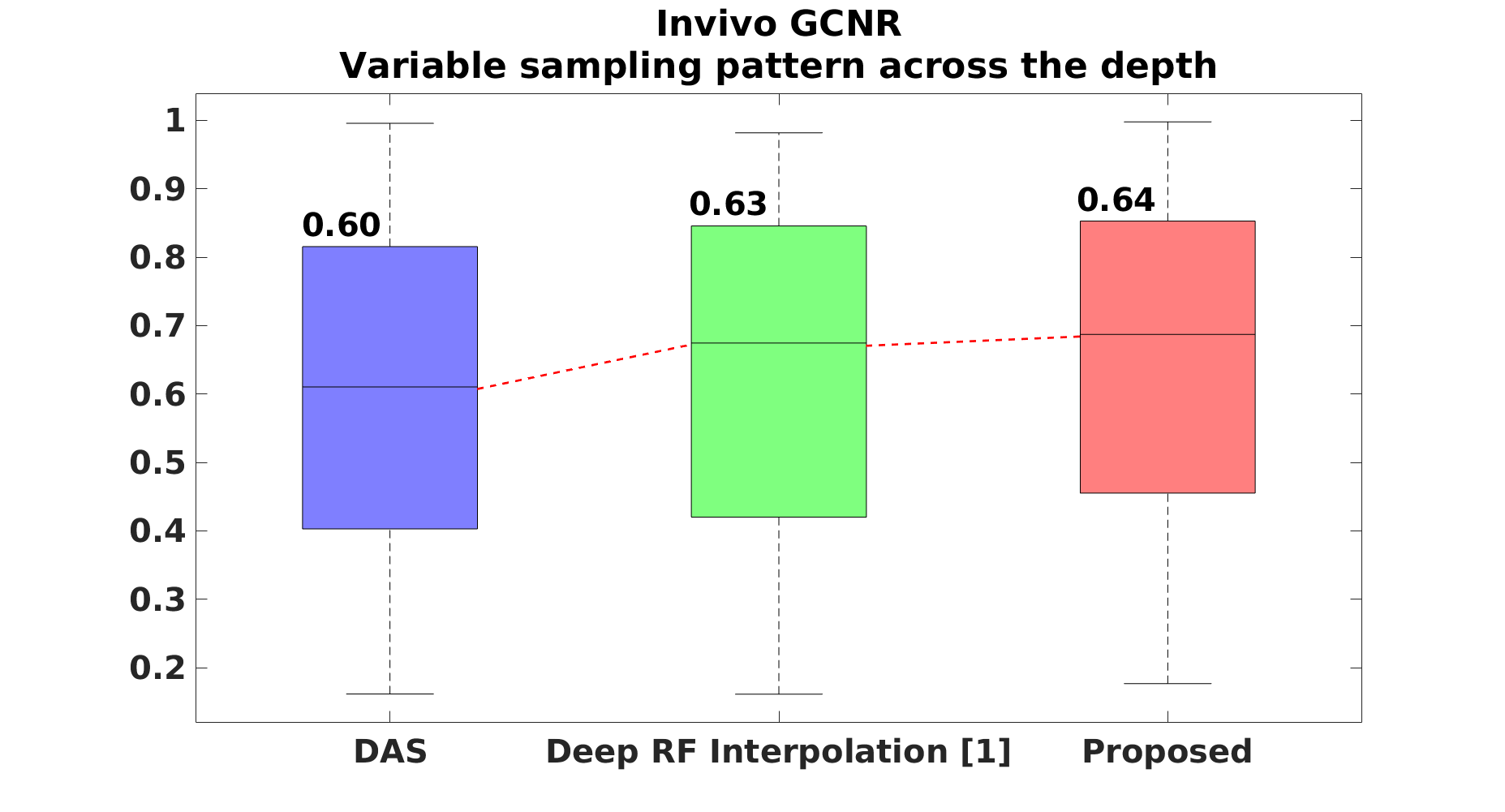}\hspace*{-0.5cm}\includegraphics[width=4cm]{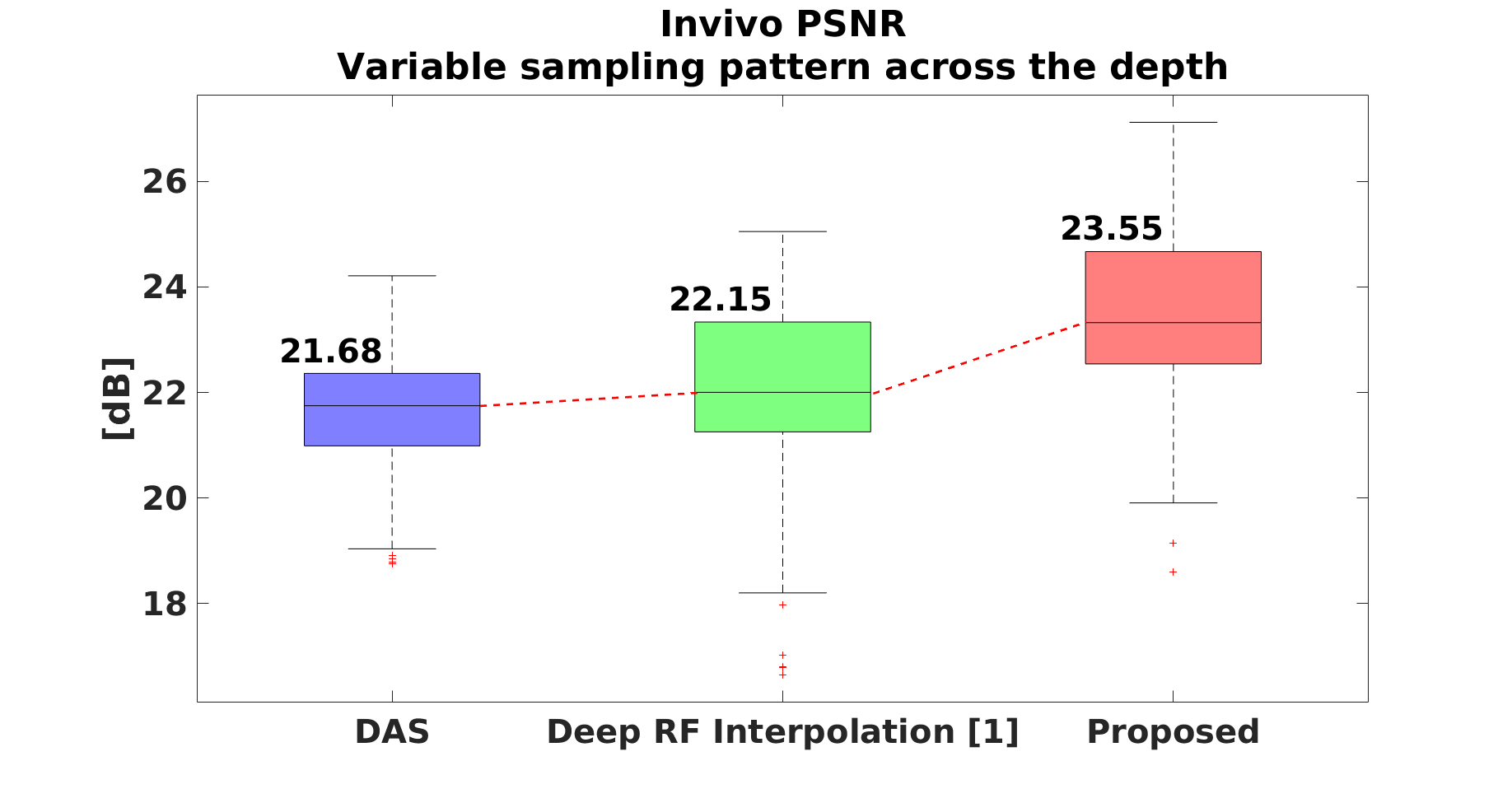}\hspace*{-0.5cm}\includegraphics[width=4cm]{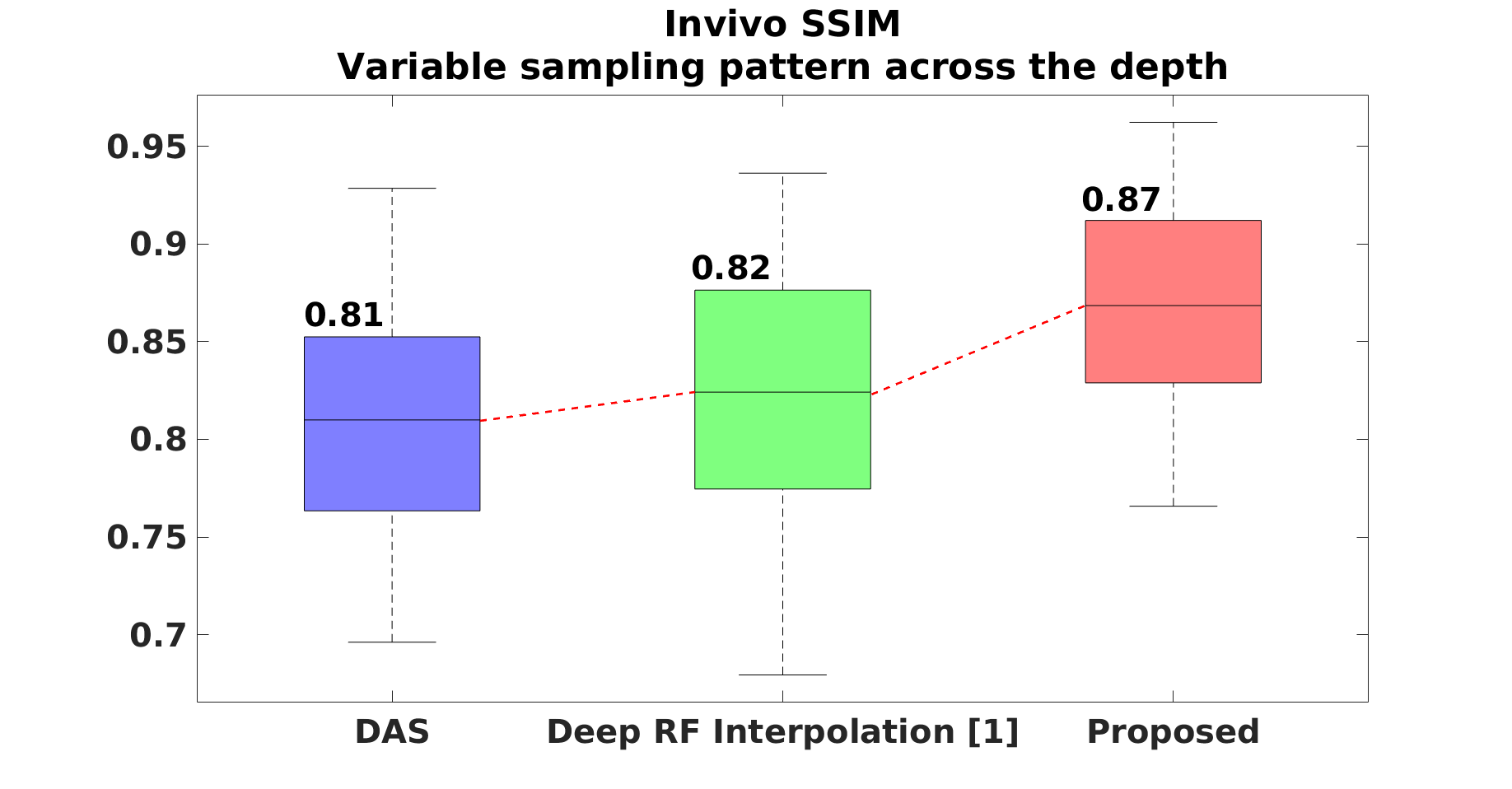}\hspace*{-0.5cm}}
	\vspace*{0.1cm}
	\caption{\footnotesize Quantitative comparison using invivo data on 4$\times$ subsampling scheme: ({first column}) CNR value distribution, ({second column}) GCNR value distribution, ({third column}) PSNR value distribution, ({fourth column}) SSIM value distribution. }
	\label{fig:results_STATS_Efficient}
\end{figure}

To verify that the proposed DeepBF still outperforms the deep RF interpolation  \cite{yoon2018efficient},
we also compared the CNR, GCNR, PSNR and SSIM distributions of reconstructed B-mode images obtained from $360$ \textit{in-vivo} test frames. Fig.~\ref{fig:results_STATS_Efficient} show distribution of aforementioned statistics on 4$\times$ sub-sampled \textit{in-vivo} dataset.  From results it can be easily seen that the proposed deep beamformer consistently outperformed the standard DAS beamformer and the Deep RF Interpolation \cite{yoon2018efficient}. Specifically, note  that on the in-vivo test dataset, the proposed network also outperform the Deep RF Interpolation \cite{yoon2018efficient}, by $0.07$ and $0.016$ units in CNR and GCNR, respectively. Whereas, in PSNR and SSIM the proposed method achieved $1.40$ dB, $0.05$ units improvement respectively. 
Fig \ref{fig:results_view_Efficient}, shows the reconstruction results on 4$\times$ sub-sampled \textit{in-vivo} data using conventional DAS, Deep RF Interpolation \cite{yoon2018efficient} and the proposed DeepBF. 

In short, the novelty of this work is the end-to-end deep learning to replace the standard BF, which was never considered in \cite{yoon2018efficient}. 

%
%
%
%
%
%
%
%